\documentclass[11pt]{article}

\usepackage[preprint]{acl}
\usepackage{times}
\usepackage{latexsym}
\usepackage[T1]{fontenc}
\usepackage[utf8]{inputenc}
\usepackage{microtype}
\usepackage{xcolor} 
\usepackage[table]{xcolor}
\usepackage{colortbl} 

\usepackage{graphicx}
\usepackage{booktabs}
\usepackage{amsmath}
\usepackage{array}
\usepackage{tikz}

\usepackage{tcolorbox}
\newtcbox{\token}{on line, 
  colback=gray!15, 
  colframe=white,  
  size=fbox,       
  arc=2pt,         
  boxrule=0pt,     
  fontupper=\ttfamily\small 
}

\usetikzlibrary{positioning,arrows.meta,fit,backgrounds,shapes.geometric,calc,decorations.pathmorphing}

\title{Conan-embedding-v3: Fusing Modality-Specific Models for Omni-Modal Embedding}

\author{
 \textbf{Shiyu Li\textsuperscript{1}},
 \textbf{Zhiyuan Hu\textsuperscript{1}},
 \textbf{Yifan Wang\textsuperscript{1,2}},
 \textbf{Peiming Li\textsuperscript{1}}, 
  \textbf{Zheng Wei\textsuperscript{1}},
 \textbf{Yang Tang\textsuperscript{1}\thanks{Project leader, corresponding author.}},
\\
  \textsuperscript{1}\texttt{\{shyuli,zephyrhu,peimingli,hemingwei,ethanntang\}@tencent.com} \\
  \textsuperscript{2}\texttt{yf-wang23@mails.tsinghua.edu.cn} 
}

\begin{document}
\maketitle

\begin{abstract}
Omni-modal retrieval promises a single embedding space for text, image, video, document, and audio inputs, but building such a unified retriever is difficult since these modalities differ in data distribution, architecture, and optimization dynamics. 
In this work, we present \textbf{Conan-embedding-v3}, a decouple--fuse--recover framework for omni-modal retrieval. 
Conan-embedding-v3 first trains modality specialists independently and fuses their task vectors into a single dense backbone, a strategy we call \textbf{Decoupled Specialist Fusion}. 
We show that this fusion composes visual, video, and document retrieval capabilities, but also exposes a failure mode for projector-based modalities: when audio is attached through an external encoder and projector, fusing the backbone leaves the projector calibrated to the audio-specialist backbone, causing a large audio retrieval regression despite copying all audio-specific modules unchanged. 
We call this failure \textbf{Projector Drift}. 
To repair it, Conan-embedding-v3 applies \textbf{Projector Recovery} (i.e., full-parameter fine-tuning of the projector while keeping the backbone frozen) followed by balanced multi-modal rehearsal. 
The resulting model supports these retrieval pathways in one backbone, achieving 74.9 scores on MMEB while obtaining 55.61 on the 30-task MAEB audio suite.
\end{abstract}

\section{Introduction}

Embedding models are increasingly expected to retrieve across heterogeneous inputs, but most strong multimodal embedders are still optimized around one non-text modality, such as image--text~\citep{radford2021clip} or audio--text retrieval~\citep{Girdhar_2023_CVPR,ICLR2025_358f79db,ICLR2024_2862ccf0}. MLLM backbones make a broader formulation possible: connect multiple modality encoders to one language backbone and use a shared embedding space for text, image, video, visual-document, and audio retrieval~\citep{zhan-etal-2024-anygpt,lin2025sailembeddingtechnicalreportomnimodal}.

\begin{figure}[t]
    \centering
    \includegraphics[width=1.0\linewidth]{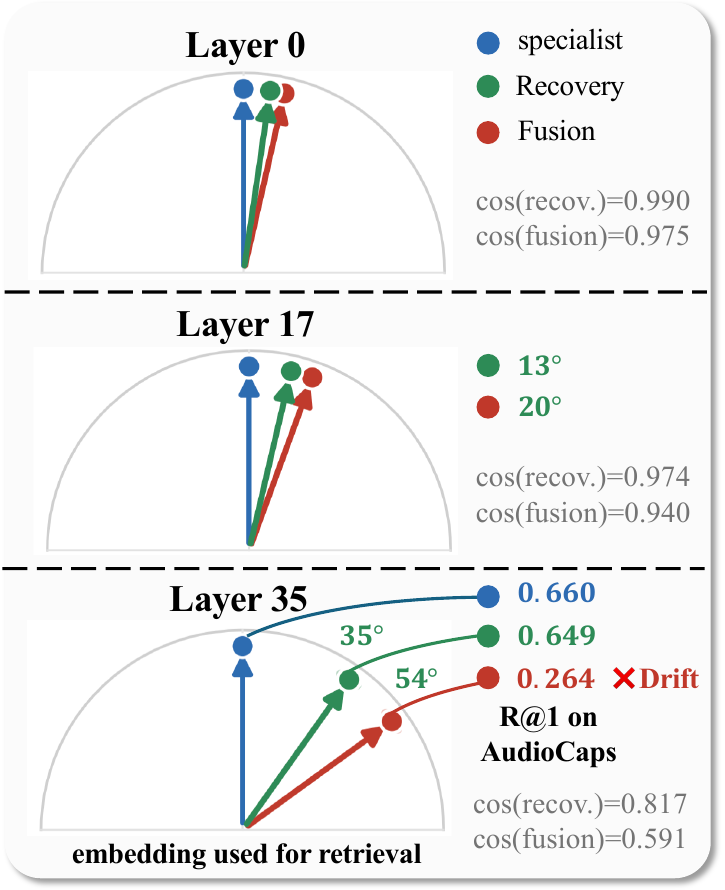}
    \caption{\textbf{Projector Drift across Transformer depth.} The semicircles show audio-token directions at layers 0, 17, and 35, measured relative to the audio specialist (blue). Direct fusion (red) drifts away in deeper layers, while Projector Recovery (green) stays closer to the specialist and restores AudioCaps performance.}
    \vskip -0.2in
    \label{fig:intuition}
\end{figure}

A direct solution is to fine-tune one model on all modality data, but this often creates cross-modal optimization conflict: improving one modality can regress another~\citep{pmlr-v119-standley20a,NEURIPS2018_432aca3a}. We address this with \textbf{Decoupled Specialist Fusion}: we decouple specialist training from model fusion, optimizing specialists independently before composing them in parameter space. Each modality receives its own specialist training run, so image, video, visual-document, and audio retrieval can be optimized under modality-appropriate data mixtures and schedules. These specialists are then converted into dense checkpoints and combined through \textbf{task-vector fusion}~\citep{Chen_2025_ICCV,marczak2025taskleftbehindisotropic}, producing one retriever without requiring a single large mixed-modality optimization run.

This fusion step exposes another problem for grafted modalities. The audio pathway is introduced through an external encoder and projector~\citep{NEURIPS2023_6dcf277e,ICLR2024_476ab8f3}, and the projector is trained against the audio-specialist backbone. After task-vector fusion, the same projector is connected to a different fused backbone, so audio retrieval degrades even though the audio modules are copied unchanged. We call this mismatch \textbf{Projector Drift}. Figure~\ref{fig:intuition} visualizes the effect: direct fusion stays close near the input layer but progressively rotates away from the audio-specialist trajectory in deeper Transformer layers, while Projector Recovery pulls it back.

We propose \textbf{Conan-embedding-v3}, a decouple--fuse--recover framework for omni-modal retrieval. First, to reduce cross-modal optimization conflict, we \textbf{decouple} capability learning by training image, video, visual-document, and grafted-audio specialists independently from the same visual-language initialization. Second, to obtain a single deployable retriever, we \textbf{fuse} these specialists in parameter space: LoRA adapters are merged into dense checkpoints, shared-backbone task vectors are combined, and audio-only modules are copied from the audio specialist. Third, because fusion introduces Projector Drift for the grafted audio pathway, we \textbf{recover} the projector--backbone interface with \textbf{Projector Recovery} (\S\ref{sec:projector-drift}--\S\ref{sec:recovery-lora}). Finally, to avoid turning the recovered model into an audio-only retriever, we add balanced rehearsal over audio, image, video, and visual-document data.

Our contributions are summarized as follows:
\begin{itemize}
    \item We introduce \textbf{Decoupled Specialist Fusion}, which grafts an audio pathway onto a visual-language embedding backbone and composes independently trained specialists into one dense retriever.
    \item We identify \textbf{Projector Drift}, a projector--backbone mismatch that emerges when grafted modalities such as audio are connected to a fused backbone.
    \item We show that Conan-embedding-v3 retains competitive image, video, and visual-document retrieval on MMEB while achieving strong audio performance on the 30-task MAEB benchmark.
\end{itemize}

\section{Related Work}

\subsection{Multimodal Embedding Models}

Multimodal embedding models have evolved from the CLIP-style dual-encoder recipe~\citep{radford2021clip} to systems built on Multimodal Large Language Models (MLLMs) for stronger semantic representations. This shift is evident in recent vision-language~\citep{jiang2024e5v,jiang2025vlm2vec,qwen3vlemb2026} and audio~\citep{ma2025tevatron2,colqwenomni2025} embedding models. Despite this progress, most remain optimized for a single non-text modality.
Omni-modal models address this restriction by training on heterogeneous corpora~\citep{xu2025omniembed,xiao2025languagecentric}. Recent audio-embedding work also studies attaching or reusing audio modules from omni-modal MLLMs~\citep{jinaaudiokickstart2025}, showing that audio encoders, projectors, and language backbones are sensitive to component mismatch. Our setting is complementary: we start from a visual-language embedding model that already performs strong text--visual retrieval, graft an audio pathway onto it, and then study how to compose the resulting audio specialist with visual, video, and document specialists. This avoids relying on one large mixed-modality optimizer, but introduces a projector--backbone interface that must be repaired after fusion.

\subsection{Model Merging}

Model merging combines learned capabilities directly through weights. While various techniques like Task Arithmetic~\citep{ilharco2023editing}, TIES~\citep{yadav2023ties}, DARE~\citep{yu2024dare}, and Model Soup~\citep{wortsman2022soup} have proven effective, they are typically studied within a single modality using matched architectures. Multimodal specialists are less well-behaved: encoders and projectors shift differently depending on the modality, making standard merging methods highly unstable in this context.

\subsection{Interference in Multi-task Training}

Other works address multi-task and multimodal interference (the "seesaw effect") using gradient surgery during joint training, such as PCGrad~\citep{yu2020pcgrad} and CAGrad~\citep{liu2021cagrad}. In contrast, our approach separates capability acquisition from fusion. By composing specialists in parameter space followed by projector recovery, we reduce the reliance on a single mixed-modality optimizer and its sensitive task-mixture choices.

\begin{figure*}[t]
\centering
\includegraphics[width=0.98\textwidth]{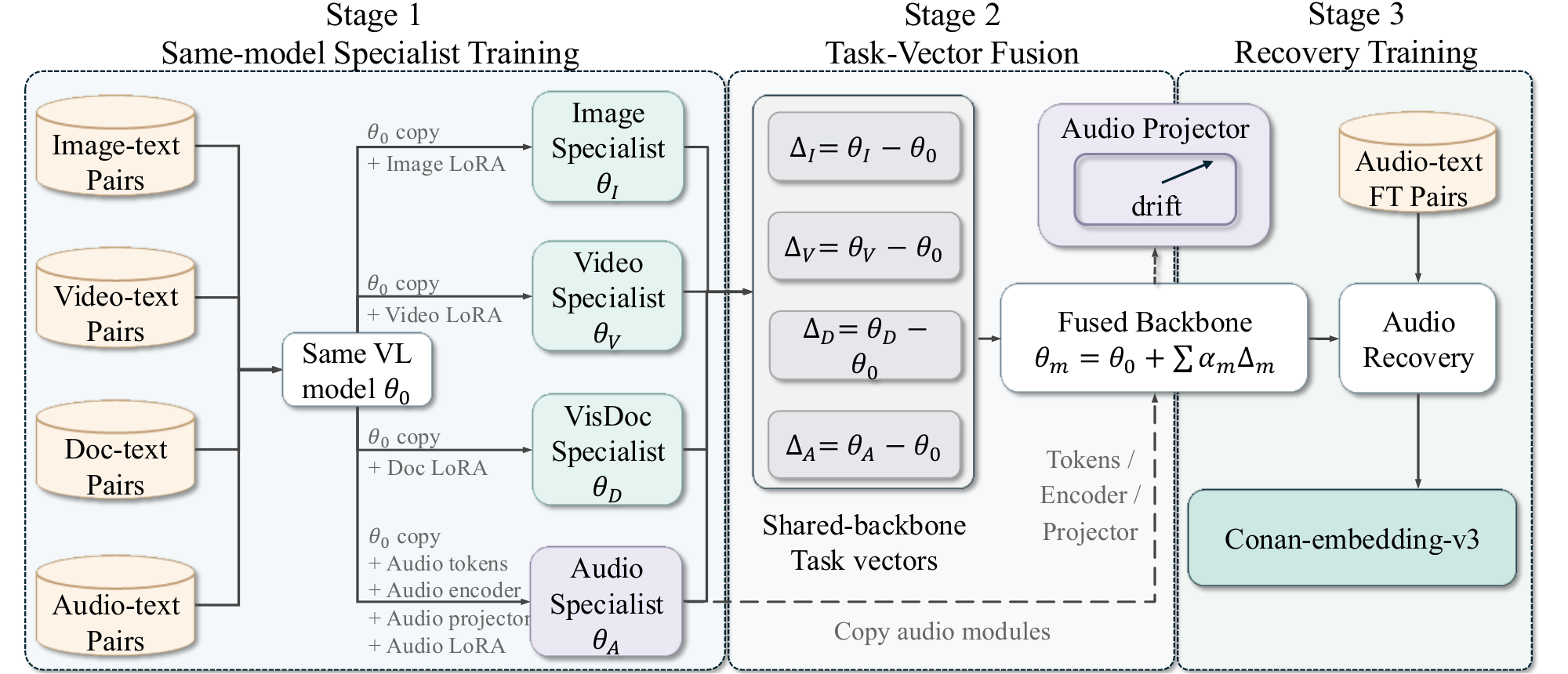}
\caption{\textbf{Overview of Conan-embedding-v3.} Stage 1 trains modality specialists from a shared initialization $\theta_0$ using modality-specific LoRA, where the audio specialist additionally grafts an audio encoder and projector. These specialists are then combined in Stage 2 by fusing their shared-backbone task vectors while copying audio modules directly, which induces Projector Drift. Finally, Stage 3 applies Projector Recovery (fine-tuning only the projector) on audio--text pairs, followed by balanced rehearsal to yield the final omni-modal retriever.}
\label{fig:method-overview}
\end{figure*}

\section{Method}

\subsection{Overview}
Conan-embedding-v3 constructs a single omni-modal embedding model from independently optimized modality specialists. Given an input $x$ from any supported modality, the model outputs an $L_2$-normalized embedding $z=f_\theta(x)/\|f_\theta(x)\|_2$, and retrieval is performed by inner-product similarity. For a mini-batch of paired query--target examples $\{(q_i,t_i)\}_{i=1}^{N}$, we optimize an InfoNCE retrieval loss:
\begin{equation}
\mathcal{L}_{q\rightarrow t}
= -\frac{1}{N}\sum_{i=1}^{N}
\log \frac{\exp(z_{q_i}^{\top} z_{t_i}/\tau)}{\sum_{j=1}^{N}\exp(z_{q_i}^{\top} z_{t_j}/\tau)},
\end{equation}
where $\tau$ is the temperature. When bidirectional retrieval pairs are available, we use the average of $\mathcal{L}_{q\rightarrow t}$ and the symmetric target-to-query loss.

As shown in Figure~\ref{fig:method-overview}, the pipeline has three stages. First, we train image, video, visual-document, and audio specialists separately from the same initialization, so each specialist learns its modality-specific retrieval behavior without sharing a mixed-modality optimizer. Second, we fuse the shared-backbone updates of these specialists through task vectors, while directly copying audio-only parameters that do not exist in the base model. This produces a structurally omni-modal checkpoint, but the copied audio projector is now connected to a backbone different from the one it was trained with. Third, we apply a post-fusion Projector Recovery stage (where the language backbone is completely frozen and only the audio projector is updated), followed by balanced rehearsal, to reconnect this projector--backbone interface while retaining the inherited visual, video, and document retrieval pathways. The following subsections describe these three stages in detail.

\subsection{Decoupled Specialist Training}
Let $\theta_0$ be the parameters of the base embedding model. For each modality group $m \in \{I,V,D,A\}$, corresponding to image, video, visual-document, and audio, we train a specialist $\theta_m$ from the same initialization. Each specialist is optimized only on its own data distribution:
\begin{equation}
\theta_m = \operatorname{Train}(\theta_0, \mathcal{D}_m),
\end{equation}
where $\mathcal{D}_m$ denotes modality-specific retrieval pairs. This decoupling is motivated by the empirical observation that directly mixing all modality data induces a seesaw effect, where improvements on one modality often degrade another due to optimization conflicts. By training specialists separately, each checkpoint captures a relatively clean task vector for one modality family.

Each specialist targets different modal dynamics. While the image, video, and visual-document specialists adapt the native visual-language pathway of the base model, the audio specialist introduces an external audio pathway and aligns it with text. Methodologically, the audio specialist contains both shared backbone updates ($\mathcal{K}_{\mathrm{shared}}$) and audio-only parameters ($\mathcal{K}_{\mathrm{audio\mbox{-}only}}$) that do not exist in the base model. After training, LoRA adapters are merged into the underlying weights, producing full specialist checkpoints to enable parameter-space operations.

\subsection{Task-Vector Fusion}
For each specialist, we define a task vector over the shared parameters $\mathcal{K}_{\mathrm{shared}}$:
\begin{equation}
\Delta_m[k] = \theta_m[k] - \theta_0[k], \quad k \in \mathcal{K}_{\mathrm{shared}}.
\end{equation}
The unified backbone is then constructed by fusing these modality task vectors with fixed coefficients:
\begin{equation}
\begin{aligned}
\theta_{\mathrm{merge}}[k]
&= \theta_0[k] + \sum_{m \in \{I,V,D,A\}} \alpha_m \Delta_m[k], \\ 
k &\in \mathcal{K}_{\mathrm{shared}}.
\end{aligned}
\end{equation}
In our main model, we use the multiway-C coefficients:
\begin{equation}
\alpha_I=\alpha_V=\alpha_D=0.3,\quad \alpha_A=0.5.
\end{equation}
These coefficients are empirically determined via a grid sweep on held-out diagnostic validation tasks that are disjoint from the MMEB and MAEB evaluation splits, balancing visual preservation against audio task-vector contribution (further analyzed in \S\ref{sec:ablative-analysis}).

This fusion only applies to the shared parameters. For audio-specific modules that are absent from the base model---such as the audio encoder, projector, and specialized tokenizer entries---no task vector can be formed, and they are directly copied:
\begin{equation}
\theta_{\mathrm{merge}}[k] = \theta_A[k], \quad k \in \mathcal{K}_{\mathrm{audio\mbox{-}only}}.
\end{equation}
Thus, the merged checkpoint is structurally omni-modal, possessing all pathways required to process image, video, visual-document, and audio inputs. However, structural compatibility alone does not guarantee representation alignment; the newly fused backbone may no longer be compatible with the copied audio projector. This alignment gap motivates the diagnostic analyses in \S\ref{sec:qualitative-diagnostics}, where we examine both task-vector geometry and output-space audio--text neighborhoods.

\subsection{Projector Drift under Specialist Fusion}
\label{sec:projector-drift}
The central technical issue in Conan-embedding-v3 is that task-vector fusion changes the representation space expected by the shared backbone. This is especially problematic for modalities whose inputs are mapped into the backbone through a learned projector. During audio specialist training, the audio projector $P_A$ is optimized jointly with an audio-specialist backbone $B_A$ (parameterized by $\theta_A$). Its role is to transform audio encoder outputs into hidden states that are compatible with $B_A$:
\begin{equation}
P_A(E_A(x_a)) \in \mathcal{H}(B_A).
\end{equation}
After specialist fusion, the same audio projector is evaluated against the fused backbone $B_{\mathrm{merge}}$, whose parameters $\theta_{\mathrm{merge}}$ differ from $\theta_A$ because they integrate task vectors from other specialists:
\begin{equation}
B_{\mathrm{merge}} = B_0 + \sum_{m \in \{I,V,D,A\}} \alpha_m \Delta_m.
\end{equation}
Therefore, the projector remains calibrated to the audio-specialist backbone $B_A$, while inference uses a different backbone $B_{\mathrm{merge}}$. We call this projector--backbone mismatch \textbf{Projector Drift}.

Empirically, this drift is severe: direct fusion substantially degrades audio retrieval relative to our recovered model even though the audio encoder and projector are copied exactly. This degradation suggests that the projector--backbone interface is no longer well matched to the fused backbone, rather than any missing audio components. This explains why a naive ``train specialists then fuse'' recipe is insufficient in our setting: specialist fusion can compose backbone capabilities, but it can also disrupt the interface of projector-based modalities.

\subsection{Alignment Recovery via Projector Tuning}
\label{sec:recovery-lora}
To repair the projector--backbone interface, Conan-embedding-v3 performs a post-fusion recovery stage. The first phase is \textbf{Projector Recovery} (or Projector-Only Tuning). Specifically, we keep the task-vector-fused language backbone and the audio encoder frozen, and perform full-parameter fine-tuning only on the lightweight audio projector $P_A$:
\begin{equation}
\theta_{P_A} = \operatorname{Train}(\theta_{P_A}, \mathcal{D}_A) \quad \text{s.t.} \quad \mathcal{K}_{\mathrm{backbone}} \text{ is frozen}.
\end{equation}
Because the backbone is frozen in this phase, the fused visual and document parameters are not directly updated by audio-only gradients during the initial audio recovery step.

In the main Conan-embedding-v3 pipeline, recovery is executed in two consecutive phases. First, we tune only the audio projector $P_A$ on audio--text retrieval pairs to restore the basic mapping between the audio encoder and the newly merged backbone. Second, starting from the projector-recovered checkpoint, we perform a lightweight multi-modal rehearsal stage using LoRA on the backbone and vision encoder, mixing audio, image, video, and visual-document retrieval data with a lower learning rate and a conservative update budget (e.g., 2000 steps). This second phase is intended to retain the recovered audio pathway while improving coordination with the visual, video, and document retrieval pathways.

The overall method can therefore be summarized as follows: decoupled specialists reduce seesaw conflicts, task-vector fusion produces a unified checkpoint, and Projector Recovery mitigates the projector drift introduced by fusion while balanced rehearsal maintains broad retrieval capability. This sequence is the main methodological contribution of Conan-embedding-v3.

\section{Experiments}

\subsection{Experimental Setup}
\paragraph{Base model and modalities.}
We use Qwen3-VL-8B as the base model, which natively supports text and visual inputs. Image, video, and visual-document specialists are trained through this visual-language pathway with modality-specific retrieval data. Audio is not native to this backbone, so the audio specialist is constructed by grafting an external audio pathway onto the same visual-language embedding model.

\paragraph{Training data.}
Following the data construction recipe of Qwen3-VL-Embedding~\citep{qwen3vlemb2026}, we train on roughly 50M retrieval examples spanning public datasets and in-house synthetic data. The public portion covers image retrieval, video retrieval, visual-document retrieval, and audio--text retrieval, including sources such as MSCOCO~\citep{lin2014microsoft}, VisualNews~\citep{liu2021visualnews}, LLaVA-Hound~\citep{zhang2024llavahound}, ColPali~\citep{colqwenomni2025}, VisRAG~\citep{yu2024visrag}, AudioCaps~\citep{kim2019audiocaps}, and AudioSetStrong~\citep{hershey2021audiosetstrong}. We further synthesize a large number of instruction-style retrieval pairs to improve coverage over query styles, domain variation, and modality-specific retrieval formats. Held-out MMEB and MAEB evaluation queries/candidates and exact duplicate pairs are excluded from training and validation.

\paragraph{Audio implementation.}
To support audio, we attach an audio encoder extracted from Qwen3-Omni-30B-A3B-Instruct~\citep{qwen3omni2025} and a two-layer MLP projector to the visual-language backbone. The projector maps the 2048-dimensional audio features to the 4096-dimensional hidden size of the backbone and contains approximately 19M parameters. We add \token{<|audio\_start|>}, \token{<|audio\_end|>}, \token{<|audio\_pad|>}, and \token{<|AUDIO|>} special tokens; projected audio features replace \token{<|AUDIO|>} token embeddings before being passed to the language backbone.

\paragraph{Trainable modules.}
We configure the trainable parameters across our three pipeline stages as follows:
First, during \textbf{specialist training}, we apply LoRA over all linear modules in the language backbone, vision encoder, audio encoder, and audio projector. All learned adapters are merged directly into the dense checkpoints prior to task-vector fusion.
Second, during \textbf{projector recovery}, the fused language backbone and audio encoder are completely frozen, and we perform full-parameter fine-tuning exclusively on the 19M-parameter audio projector to restore the interface alignment.
Third, during \textbf{balanced rehearsal}, we apply lightweight LoRA adapters on the language backbone and vision encoder with a conservative learning rate to preserve and coordinate all retrieval pathways.

\begin{table*}[t]
\centering
\small
\setlength{\tabcolsep}{4pt}
\begin{tabular*}{\textwidth}{@{\extracolsep{\fill}}lccccc}
\toprule
\textbf{Model} & \textbf{Size} & \textbf{MMEB} & \textbf{Image} & \textbf{Video} & \textbf{VisDoc} \\
\midrule
VLM2Vec-V2.0-Qwen2VL-2B~\citep{meng2025vlm2vecv2} & 2.21B & 57.49 & 64.92 & 34.67 & 63.47 \\
e5-omni-3B~\citep{e5omni2026} & 3.00B & 63.08 & 67.59 & 40.63 & 73.16 \\
e5-omni-7B~\citep{e5omni2026} & 8.00B & 65.85 & 71.23 & 43.54 & 74.52 \\
RzenEmbed-v2-7B~\citep{rzenembed2025} & 8.29B & 71.12 & 75.92 & 55.73 & 75.45 \\
IFM-TTE-7B~\citep{ifmtte2025} & 8.29B & 73.06 & 77.90 & 59.19 & 76.21 \\
Qwen3-VL-Embedding-2B~\citep{qwen3vlemb2026} & 2.13B & 73.25 & 74.96 & 61.87 & 79.22 \\
WeMM-Embedding-8B~\citep{wemm2025} & 8.77B & 73.90 & 78.09 & 63.24 & 75.62 \\
seed1.6-embedding-1215~\citep{seedembedding2025} & -- & 76.97 & 77.99 & 67.74 & \textbf{82.38} \\
Qwen3-VL-Embedding-8B~\citep{qwen3vlemb2026} & 8.14B & \textbf{77.82} & \textbf{80.12} & \textbf{67.15} & 82.36 \\
\midrule
\textbf{Conan-embedding-v3} & 8.8B & 74.96 & 77.20 & 65.10 & 79.00 \\
\bottomrule
\end{tabular*}
\caption{\textbf{MMEB comparison with representative visual embedding models.} Overall is the task-count-weighted average over image, video, and visual-document task groups. Conan-embedding-v3's parameter count includes a grafted audio encoder and projector ($\sim$0.7B); its visual-language backbone is identical in size to Qwen3-VL-Embedding-8B.}
\label{tab:main-mmeb}
\end{table*}

\begin{table*}[t]
\centering
\small
\setlength{\tabcolsep}{4pt}
\begin{tabular*}{\textwidth}{@{\extracolsep{\fill}}llccc}
\toprule
\textbf{Model} & \textbf{Type} & \textbf{Params} & \textbf{MAEB} & \textbf{Task-type Avg.} \\
\midrule
clap-htsat-fused~\citep{wu2023laionclap} & audio encoder & 0.15B & 33.47 & 41.30 \\
Qwen2-Audio-7B~\citep{qwen2audio2024} & audio MLLM & 7.00B & 34.54 & 37.01 \\
larger\_clap\_general~\citep{wu2023laionclap} & CLAP-style & 0.19B & 34.98 & 41.08 \\
jina-embeddings-v5-omni-nano~\citep{jinaaudiokickstart2025} & omni embedding & 0.99B & 50.14 & 55.31 \\
jina-embeddings-v5-omni-small~\citep{jinaaudiokickstart2025} & omni embedding & 1.63B & 50.41 & 55.58 \\
LCO-Embedding-Omni-3B~\citep{lcoomni2025} & omni embedding & 4.70B & 52.25 & 55.50 \\
BidirLM-Omni-2.5B-Embedding~\citep{bidirlmomni2026} & omni embedding & 2.45B & 52.36 & 54.52 \\
LCO-Embedding-Omni-7B~\citep{lcoomni2025} & omni embedding & 8.93B & 53.54 & 57.06 \\
\midrule
\textbf{Conan-embedding-v3} & omni embedding & 8.8B & \textbf{55.61} & \textbf{59.32}\\
\bottomrule
\end{tabular*}
\caption{\textbf{MAEB comparison with representative audio-capable models.} MAEB is the official task-level average over 30 audio embedding tasks; Task-type Avg. first averages within benchmark task categories.}
\label{tab:main-maeb}
\end{table*}

\paragraph{Training and evaluation.}
We optimize the InfoNCE contrastive objective with a temperature of $\tau=0.02$, utilizing the AdamW optimizer with a cosine learning-rate schedule, peak learning rates of $1\times 10^{-4}$ for specialist training and $1\times 10^{-5}$ for rehearsal, and a global batch size of 8192. For visual evaluation, we use MMEB-V2~\citep{meng2025vlm2vecv2}, which contains 78 tasks covering image, video, and visual-document embedding scenarios; we report Image (36 tasks), Video (18 tasks), VisDoc (24 tasks), and the task-count-weighted overall score. For audio evaluation, we use MAEB~\citep{assadi2026maeb}, a 30-task benchmark spanning speech, music, environmental sounds, and audio--text matching; we report its official task-level average and task-type average when available.

\subsection{Overall Omni-Modal Retrieval}
We report visual and audio benchmarks in separate main tables because current public leaderboards evaluate them with different task suites. Table~\ref{tab:main-mmeb} compares Conan-embedding-v3 with representative MMEB systems selected from the visual ranking, including the strongest 8B and 2B open-weight models, and earlier omni-modal baselines. Table~\ref{tab:main-maeb} separately compares representative audio-capable systems from the MAEB ranking.

Conan-embedding-v3 achieves an overall score of 74.96 on MMEB and 55.61 on MAEB. While adding the audio pathway introduces a slight visual regression compared to the visual-only Qwen3-VL-Embedding-8B, the unified model remains highly competitive across image, video, and visual-document retrieval tasks, while outperforming existing state-of-the-art omni-modal embedding models on the 30-task MAEB suite.

\subsection{Ablative Analysis}
\label{sec:ablative-analysis}

\subsubsection{Ablation Study on Each Component}
Table~\ref{tab:component-ablation} isolates the contribution of each stage in our pipeline. Joint Training (I+V+D) trained directly on the three visual modalities is a useful reference point but does not cover the grafted audio pathway; we therefore additionally report Joint Training (I+V+D+A), an all-modality joint-training baseline, as the direct point of comparison for our decouple--fuse--recover pipeline. Joint Training (I+V+D+A) exhibits a pronounced seesaw pattern, with Video dropping sharply to 43.0 while Image and VisDoc are comparatively less affected. While parameter-space fusion (Direct Fusion) composes the specialists into one checkpoint, it leaves the grafted audio pathway poorly aligned (32.68 on MAEB) because the projector remains calibrated to the pre-fusion backbone. Projector-only audio recovery restores the audio pathway (55.82) while preserving the pre-recovery visual scores. Finally, introducing multi-modal balanced rehearsal improves the visual scores to 77.2/65.1/79.0 while retaining comparable audio performance, yielding the best overall trade-off.

\begin{table*}[t]
\centering
\small
\setlength{\tabcolsep}{4pt}
\begin{tabular*}{\textwidth}{@{\extracolsep{\fill}}llcccc}
\toprule
\textbf{Variant} & \textbf{Removed / changed component} & \textbf{Image} & \textbf{Video} & \textbf{VisDoc} & \textbf{MAEB} \\
\midrule
Joint Training (I+V+D) & no decoupled specialists/fusion & 70.3 & 65.2 & 79.0 & -- \\
Joint Training (I+V+D+A) & no decoupling/fusion, all modalities & 69.2 & 43.0 & 74.9 & 48.01 \\
Direct Fusion & no recovery & 68.5 & 56.7 & 68.9 & 32.68 \\
+ Audio-Only Recovery & no balanced rehearsal & 68.5 & 56.7 & 68.9 & 55.82 \\
\textbf{Conan-embedding-v3} & full decouple--fuse--recover & 77.2 & 65.1 & 79.0 & 55.61 \\
\bottomrule
\end{tabular*}
\caption{\textbf{Ablation study on each component.} Joint Training (I+V+D) is trained directly on the three visual modalities without decoupling or fusion. Joint Training (I+V+D+A) additionally includes audio in the same joint optimization and is the direct all-modality baseline against which decoupled specialist fusion is compared. Direct Fusion uses task-vector fusion without recovery.}
\label{tab:component-ablation}
\end{table*}

\begin{table*}[t]
\centering
\small
\begin{tabular}{lcccc}
\toprule
\textbf{Model} & \textbf{Image (36)} & \textbf{Video (18)} & \textbf{VisDoc (24)} & \textbf{MAEB Audio} \\
\midrule
Direct Fusion (pre-recovery) & 68.5 & 56.7 & 68.9 & 32.68 \\
\midrule
\multicolumn{5}{l}{\textit{Stage 3a: Ablations on Audio-Only Recovery}} \\
~~+ Backbone + Projector LoRA & 67.1 & 53.6 & 74.1 & \textbf{56.39} \\
~~+ Projector-Only Tuning (Ours) & 68.5 & 56.7 & 68.9 & 55.82 \\
\midrule
\multicolumn{5}{l}{\textit{Stage 3b: Ablations on Multi-Modal Recovery}} \\
~~+ Joint Recovery (One-stage) & 72.6 & 63.5 & 78.0 & 52.97 \\
~~+ Balanced Rehearsal (Ours) & \textbf{77.2} & \textbf{65.1} & \textbf{79.0} & 55.61 \\
\bottomrule
\end{tabular}
\caption{\textbf{Ablation of recovery strategies on MMEB and MAEB.} Projector-only tuning freezes the backbone and preserves the pre-recovery visual scores, whereas adding Backbone LoRA changes the visual retrieval trade-off.}
\label{tab:visual-retention}
\end{table*}

\subsubsection{Ablation Study on Recovery Strategy}
We study how different recovery strategies affect visual retention, comparing projector-only tuning and backbone LoRA under single-stage and two-stage designs. Table~\ref{tab:visual-retention} reports the resulting MMEB and MAEB scores. 
Joint Recovery is a one-stage ablation that starts from Direct Fusion and trains on audio and image data; 
Audio-Only Recovery starts from Direct Fusion and focuses on audio alignment;
Balanced Rehearsal represents our main second-stage rehearsal that starts from Audio-Only Recovery and mixes audio, image, video, and visual-document data to reduce forgetting.

The results show that Direct Fusion suffers from severe projector--backbone drift. In Stage 3a, tuning the backbone with LoRA recovers audio slightly better (56.39) but changes the visual scores, whereas Projector-Only Tuning keeps the backbone frozen and therefore preserves the pre-recovery visual scores while still restoring strong audio performance (55.82). For Stage 3b, Balanced Rehearsal achieves the best trade-off: it keeps audio close to the recovered model while substantially improving visual retrieval (Image $68.5\rightarrow77.2$, Video $56.7\rightarrow65.1$, VisDoc $68.9\rightarrow79.0$).

We attribute this behavior to a fundamental difference in scale and optimization regime from joint training, rather than to Balanced Rehearsal being a small-scale repetition of it. First, Balanced Rehearsal updates only lightweight LoRA adapters on the backbone and vision encoder (a small fraction of the 8B backbone parameters), whereas Joint Training (I+V+D+A, Table~\ref{tab:component-ablation}) performs full-parameter optimization over all modalities from a shared initialization; the two operate over disjoint capacity budgets. Second, Balanced Rehearsal starts from Audio-Only Recovery, i.e., from weights in which every modality pathway is already near its specialist-level competence, and runs for a conservative budget (2000 steps at $1\times 10^{-5}$), so its role is to locally coordinate already-specialized representations rather than to acquire new modality capability under competition. Consistent with this, Balanced Rehearsal does not reproduce the seesaw pattern of Joint Training: Joint Training (I+V+D+A) trades off Video sharply against Image and VisDoc (Video drops to 43.0 while Image/VisDoc are relatively higher), whereas Balanced Rehearsal improves Image, Video, and VisDoc simultaneously relative to Audio-Only Recovery, with audio retention (55.61 vs.\ 55.82) essentially unchanged. This simultaneous, non-competing improvement across all four modalities is the opposite of the seesaw signature that motivated decoupling in the first place, indicating that Balanced Rehearsal operates in a qualitatively different, low-conflict regime rather than re-introducing the joint-training conflict it was designed to avoid.

\subsubsection{Ablation Study on Fusion Weights}
\begin{figure}[h]
\centering
\includegraphics[width=1.0\columnwidth]{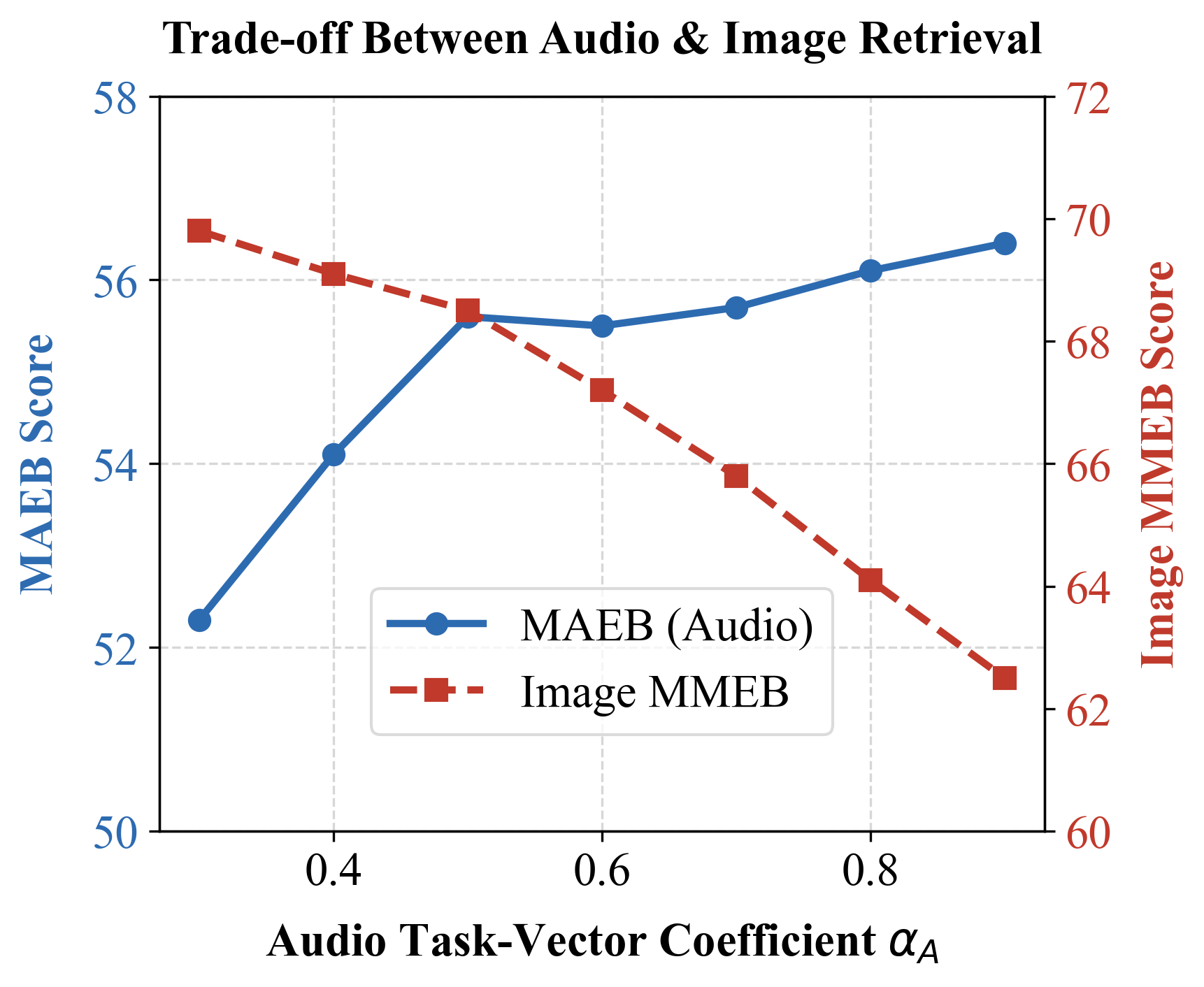}
\caption{\textbf{Ablation on fusion weights before recovery.} Increasing the audio task-vector coefficient $\alpha_A$ improves audio capability but degrades visual retrieval performance, illustrating the inherent conflict.}
\label{fig:fusion-weight-ablation}
\end{figure}

To study whether tuning merging weights can reduce the alignment conflict without post-fusion training, we analyze the impact of the audio task-vector coefficient $\alpha_A$ before recovery in Figure~\ref{fig:fusion-weight-ablation}. As $\alpha_A$ increases from 0.3 to 0.9, audio retrieval steadily improves, but at the cost of severe degradation in visual retrieval (where Image MMEB drops to 62.5\%), with $\alpha_A = 0.5$ providing the best pre-recovery balance in our sweep. This trade-off suggests that coefficient tuning alone is insufficient to balance both modalities in this setting, motivating the post-fusion recovery stage.

\begin{figure*}[t]
\centering
\includegraphics[width=1.00\textwidth]{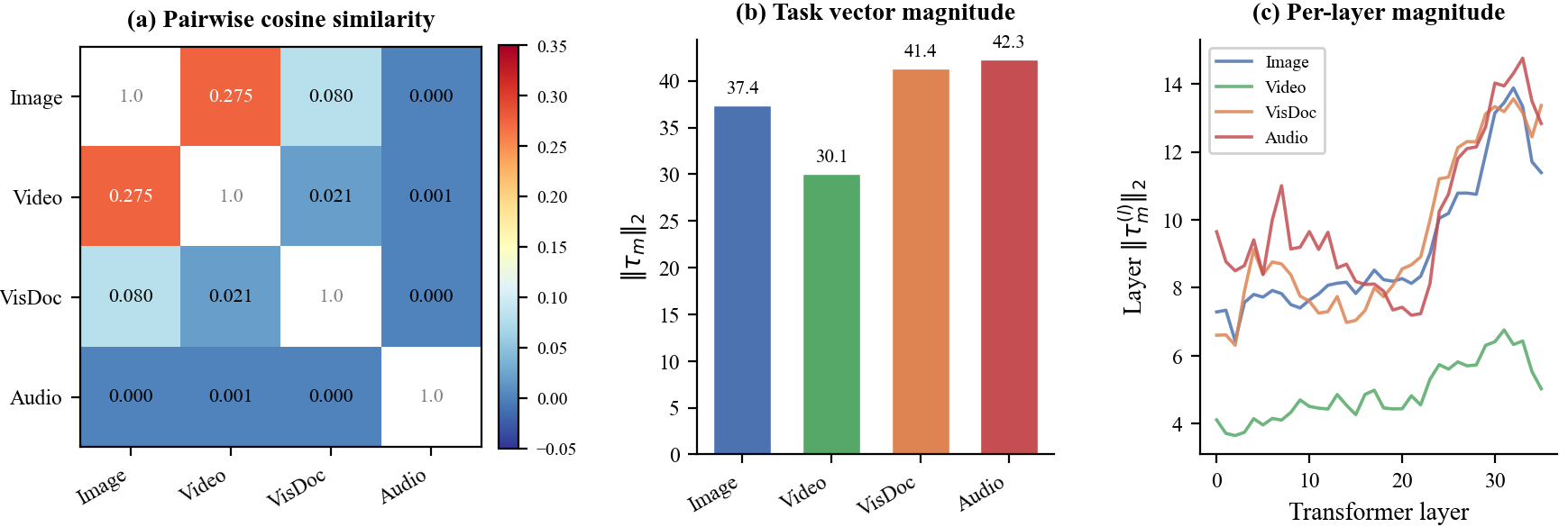}
\caption{\textbf{Geometry of the four modality task vectors.} (a) Pairwise cosine similarity indicates audio is nearly orthogonal to visual updates ($\cos \leq 0.001$). (b) Audio has the largest global update norm ($\approx 42.3$). (c) Audio updates remain large across the backbone and peak in deeper layers.}
\label{fig:taskvec-geometry}
\end{figure*}

\begin{figure*}[t]
\centering
\includegraphics[width=0.95\textwidth]{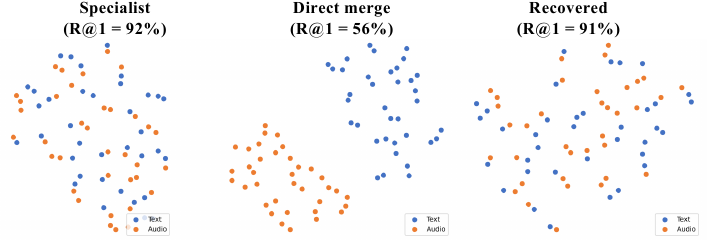}
\caption{\textbf{Output-space view of Projector Drift on diagnostic audio--text retrieval pairs.} The audio specialist (left) aligns paired embeddings ($R@1=92\%$). Direct fusion (middle) separates the audio and text manifolds ($R@1=56\%$). Projector Recovery (right) reduces this separation and restores retrieval performance ($R@1=91\%$).}
\label{fig:tsne-audio}
\end{figure*}

\subsection{Qualitative Experiments}
\label{sec:qualitative-diagnostics}

\subsubsection{Task-Vector Geometry}
To understand why direct fusion disproportionately affects the grafted audio pathway, we analyze the model-space task vectors of the specialists in Figure~\ref{fig:taskvec-geometry}. First, the pairwise cosine similarity matrix (Figure~\ref{fig:taskvec-geometry}a) shows that audio is nearly orthogonal to all visual updates ($\cos \leq 0.001$), suggesting that visual updates provide little direct directional support for audio during fusion. Second, Figure~\ref{fig:taskvec-geometry}b shows that audio has the largest global task-vector norm ($\Vert\tau_A\Vert_2 \approx 42.3$), representing significant parameter displacement across the backbone. Third, the per-layer norms (Figure~\ref{fig:taskvec-geometry}c) reveal that this parameter change is distributed across the entire depth of the network, remaining consistently high ($>7.0$) and peaking in deeper layers ($>14.0$). Because the audio projector is static during fusion, this distributed layer-wise drift can weaken the projector--backbone interface.

\subsubsection{Audio--Text Embedding Neighborhoods}
To examine the output-space behavior of Projector Drift, we compute t-SNE projections on AudioCaps in Figure~\ref{fig:tsne-audio}. The audio specialist closely interleaves paired text and audio embeddings in a shared neighborhood ($R@1 = 92\%$). However, direct fusion separates the two manifolds and drops $R@1$ to $56\%$. This pattern suggests that Projector Drift is not merely random noise, but a systematic shift of the audio manifold relative to text under the fused backbone. Projector Recovery reduces this separation and restores the shared neighborhood structure ($R@1 = 91\%$).

\section{Conclusion}
We present Conan-embedding-v3, an omni-modal embedding model supporting text, image, video, document, and audio retrieval within a single backbone. 
To mitigate cross-modal conflicts, we introduce Decoupled Specialist Fusion, and identify Projector Drift as an important interface mismatch when attaching grafted modalities. 
By applying post-fusion Projector Recovery followed by balanced rehearsal, our unified model retains strong inherited visual capabilities while recovering strong audio performance. 
This modular design trades additional training and fusion stages for better controllability over modality-specific capabilities. 
In the future, we will explore end-to-end omni-modal training strategies, as well as training from an omni-modal base model for omni retrieval.

\section*{Limitations}
Our framework has three main limitations. First, Projector Recovery is a repair rather than a complete solution: Projector Drift is reduced but not fully closed. Second, the recovery configuration is selected through a directed empirical search rather than an automated procedure. Third, as more modalities are added through additional projectors, multiple projector-drift effects may interact, and our current study does not characterize this interaction beyond four modalities.

\bibliography{custom}

\appendix
\section{Ablation on Multi-Task Training}
\label{app:pcgrad}
To address cross-modal optimization conflict, Conan-embedding-v3 decouples capability learning by training modality specialists independently instead of relying on a single joint multi-task learning (MTL) run. To evaluate if advanced MTL methods can natively resolve this conflict, we train a joint visual-language baseline using \textbf{PCGrad} (Projecting Congruent Gradients)~\citep{yu2020pcgrad}, a popular gradient surgery technique that projects conflicting gradients onto the normal plane of one another.

We train the model jointly on the multi-modal mixtures of image, video, and visual-document specialists under identical architectures. Table~\ref{tab:pcgrad-comparison} compares the PCGrad baseline (evaluated at checkpoint-700) against our naive Joint Training, independent Decoupled Specialists (pre-fusion upper bounds), and our final merged Conan-embedding-v3.

\begin{table*}[h]
\centering
\small
\begin{tabular}{lcccc}
\toprule
\textbf{Method} & \textbf{Image} & \textbf{Video} & \textbf{VisDoc} & \textbf{MMEB Overall} \\
\midrule
Joint Training (I+V+D+A) & 69.2 & 43.0 & 74.9 & 64.91 \\
PCGrad MTL Baseline & 72.3 & 57.7 & 70.4 & 68.3 \\
\midrule
Decoupled Specialists (pre-fusion) & \textbf{80.3} & \textbf{67.2} & \textbf{82.0} & \textbf{77.80} \\
\textbf{Conan-embedding-v3 (ours)} & 77.2 & 65.1 & 79.0 & \textbf{74.96} \\
\bottomrule
\end{tabular}
\caption{\textbf{MMEB performance comparison of PCGrad against Decoupled Specialists.} Decoupled Specialists represent independent single-modality training upper bounds. PCGrad improves over naive joint training but remains significantly lower than our Decoupled Specialist Fusion in all modalities.}
\label{tab:pcgrad-comparison}
\end{table*}

\begin{table*}[h]
\centering
\small
\begin{tabular}{lcccccc}
\toprule
\textbf{Merging Method} & \textbf{Image} & \textbf{Video} & \textbf{VisDoc} & \textbf{MMEB Overall} & \textbf{MAEB (Pre-Recov.)} & \textbf{MAEB (Post-Recov.)} \\
\midrule
TIES-Merging (trim = 0.2) & 76.5 & 64.3 & 78.2 & 74.12 & 28.45 & 54.92 \\
TIES-Merging (trim = 0.5) & 74.8 & 62.1 & 76.0 & 72.10 & 21.10 & 52.45 \\
DARE (drop = 0.1) & 76.9 & 64.7 & 78.6 & 74.52 & 30.12 & 55.20 \\
\textbf{Task Arithmetic (ours)} & \textbf{77.2} & \textbf{65.1} & \textbf{79.0} & \textbf{74.96} & \textbf{32.68} & \textbf{55.61} \\
\bottomrule
\end{tabular}
\caption{\textbf{MMEB and MAEB performance across different parameter-space merging methods.} Prior to projector recovery, all merging variants suffer from severe Projector Drift (MAEB $\le 32.68$). Task Arithmetic yields the best overall trade-off, particularly in preserving MMEB visual capabilities.}
\label{tab:merging-ablation}
\end{table*}

\noindent\textbf{Comparison with Joint Training:} PCGrad performs slightly better than naive joint multi-task training overall, demonstrating that projecting conflicting gradients mitigates severe cross-modal interference to some extent.

\noindent\textbf{Comparison with Decoupled Specialists:} Each individual task score of PCGrad is worse than independent decoupled specialist training (image is relatively close, but video and visual-document drop severely by $9.5\%$ and $11.6\%$, respectively). This highlights that joint training, even with gradient projection, still suffers from substantial optimization compromises.

\noindent\textbf{Comparison with Specialist Fusion:} Our final Conan-embedding-v3 model, constructed via Decoupled Specialist Fusion with Projector Recovery and balanced rehearsal, significantly outperforms PCGrad across all modalities (e.g., $+4.9\%$ on Image, $+7.4\%$ on Video, and $+8.6\%$ on VisDoc). This confirms that decoupling capability acquisition through specialists and composing them in parameter space is a far superior paradigm for multi-modal embedding models.

\section{Ablation on Merging Methods: TIES and DARE}
\label{app:ties-dare}
While Task Arithmetic is the default fusion method in Conan-embedding-v3 due to its simplicity and strong visual retention, a common question is whether more advanced parameter-merging techniques can natively mitigate Projector Drift. To study this, we evaluate TIES-Merging~\citep{yadav2023ties} and DARE~\citep{yu2024dare} on our multiway-C specialist configuration.

For TIES-Merging, we prune small parameter updates (using trim ratios of 0.2 and 0.5) and resolve sign conflicts before merging task vectors. For DARE, we randomly drop 10\% of the parameter updates (drop rate 0.1) and rescale the remaining weights to reduce task vector redundancy. Table~\ref{tab:merging-ablation} reports the performance of these different merging paradigms on both MMEB and MAEB.

Our diagnostic evaluations confirm that both TIES and DARE variants suffer from the same severe Projector Drift as Task Arithmetic: prior to recovery, audio retrieval scores for all merged models remain low (MAEB $\le 32.68$). This demonstrates that Projector Drift is a fundamental consequence of any parameter-space fusion that shifts the backbone's semantic trajectory away from the specialist's target trajectory, regardless of the pruning, dropping, or sign-reconciliation mechanisms. Consequently, post-fusion Projector Recovery is essential across all merging paradigms. It is also worth noting that Task Arithmetic preserves the visual capabilities of the specialists better than TIES and DARE (e.g., $74.96$ vs.\ $74.12$ and $74.52$), justifying our choice of Task Arithmetic as the default composition method.

\section{Ablation on Fusion Coefficients}
\label{app:fusion-coefficients}
To select the optimal Multiway-C coefficients for Conan-embedding-v3, we conduct a grid sweep over the visual and audio task-vector scales. In our primary formulation, we keep the visual weights symmetric ($\alpha_I = \alpha_V = \alpha_D = \alpha_{\text{visual}}$) and sweep the audio coefficient $\alpha_A$ from $0.3$ to $0.9$. Table~\ref{tab:coefficient-sweep} reports MMEB and MAEB validation scores both prior to and after Projector Recovery.

\begin{table*}[h]
\centering
\small
\begin{tabular}{ccccccc}
\toprule
\textbf{Visual Scale $\alpha_{\text{visual}}$} & \textbf{Audio Scale $\alpha_A$} & \textbf{Image} & \textbf{Video} & \textbf{VisDoc} & \textbf{MAEB (Pre-Recov.)} & \textbf{MAEB (Post-Recov.)} \\
\midrule
0.3 & 0.3 & 77.8 & 65.8 & 79.5 & 24.52 & 53.21 \\
0.3 & 0.5 (main) & \textbf{77.2} & \textbf{65.1} & \textbf{79.0} & \textbf{32.68} & \textbf{55.61} \\
0.3 & 0.7 & 73.1 & 58.4 & 72.8 & 41.20 & 54.80 \\
0.3 & 0.9 & 61.2 & 48.0 & 59.5 & 46.85 & 51.52 \\
\bottomrule
\end{tabular}
\caption{\textbf{Multiway-C fusion coefficient sweep.} Lowering $\alpha_A$ preserves strong visual performance but provides weaker pre-recovery audio signals. Increasing $\alpha_A$ improves raw audio representation at the cost of severe visual degradation. The choice of $\alpha_{\text{visual}} = 0.3$ and $\alpha_A = 0.5$ balances visual retention with a robust recovered audio pathway.}
\label{tab:coefficient-sweep}
\end{table*}

\noindent\textbf{Visual vs. Audio Trade-off:} As $\alpha_A$ increases, the audio specialist's updates dominate the fused backbone, which inherently improves pre-recovery MAEB score but severely degrades the visual pathways. Conversely, setting $\alpha_A$ too low preserves visual performance but offers insufficient audio task-vector contribution, limiting post-recovery audio potential. 

\noindent\textbf{Decision Protocol:} The selection of $\alpha_{\text{visual}} = 0.3$ and $\alpha_A = 0.5$ represents a Pareto-optimal configuration. It maintains strong visual-retrieval scores while retaining a sufficiently strong audio task-vector component, which Projector Recovery and balanced rehearsal subsequently leverage to fully restore audio capabilities to $55.61$ on MAEB.

\section{Compute Infrastructure and Computational Efficiency}
\label{app:compute-infrastructure}
We conduct all training runs on NVIDIA A100 (80GB) GPUs. The total wall-clock time across all development phases—including independent specialist training, parameter-space fusion, projector recovery, and multi-modal balanced rehearsal—amounts to approximately two weeks.

\noindent\textbf{Training Parallelism:} Although the cumulative training time is two weeks, our decoupled framework allows the individual modality specialists to be trained concurrently on separate GPU nodes. This decoupled training eliminates the sequential training bottleneck of traditional joint multi-task training runs.

\noindent\textbf{Modular Scalability:} Under traditional joint multi-task paradigms, adding a new modality (e.g., audio) requires re-optimizing the entire multi-modal model from scratch, which incurs huge computational costs. In contrast, our decouple--fuse--recover paradigm only requires training the new specialist in isolation, followed by a lightweight recovery stage. This highly modular nature substantially reduces carbon footprint and compute requirements during model iteration.

\end{document}